# Common Envelopes, Gamma Rays, and Sudden Spectral Changes of Novae

Robert Williams
*Space Telescope Science Institute, 3700 San Martin Dr., Baltimore, MD  21218*
*Department of Astronomy & Astrophysics, Univ. of California/Santa Cruz, 1156 High Street, Santa Cruz, CA 95064*

Russell Ryan
*Space Telescope Science institute, 3700 San Martin Dr., Baltimore, MD  21218*

Richard Rudy
*Kookoosint Scientific, 1530 Calle Portada, Camarillo, CA   93010*

**Abstract**:  A common envelope (CE) is proposed as the origin of the early postoutburst spectra of many novae.  A simple model is proposed to explain the properties of the CE based on the emission line strengths and an assumed density distribution.  Rapid changes in the spectrum during postoutburst decline are suggested as possible evidence for a CE.  Time-resolved spectra from the ARAS group show sudden spectral shifts that are correlated with detected $\gamma$-ray emission, suggestive of its possible origin on the WD that produces a change in condition within the CE.  Episodic mass loss, formation of *thea* transient heavy element absorption systems, and dissipation of the CE may be triggered by $\gamma$-ray emission.

## 1.  Introduction

   Early studies of the spectra of galactic novae were undertaken by Beals (1931), McLaughlin (1942), and Payne-Gaposchkin (1957), each of whom described the spectra in terms of distinct phases through which the spectra pass after outburst.  Subsequently, atomic physics processes were incorporated into radiative transfer calculations to predict the spectra from an expanding gas, based on input parameters that could be obtained from photometry and time-resolved spectral observations (Sobolev 1960, 1969; Castor 1972).  Because close mass transfer binaries are spatially unresolved with current instruments, one of the parameters most difficult to specify for outburst ejecta is their geometry.  It is feasible to treat complex geometry when using Monte Carlo techniques to predict the resulting spectrum from assumed input parameters, however the problem is made more tractable by assuming a regular structure that can be solved analytically.

   Studies of individual novae have been successful in representing the ejecta by evolving winds (cf. Hauschildt et al. 1992, 1995) and discrete shells of gas ejected by the outburst, whose changing properties from expansion drive the evolution of its spectrum (Shore 2012; Shore et al. 2016; Hachisu & Kato 2022).  Analysis of time-series high resolution spectra of novae during postoutburst decline, Aydi et al. (2020a, 2020b) has demonstrated that discrete episodes of ejecta that collide due to their different velocities explain key features of observed spectra.  The above studies have all been successful in characterizing the spectra of selected novae.

   An additional geometrical configuration that merits consideration in modeling the formation of novae spectra is that of a common envelope.  Independent hydrodynamical studies of the outburst have demonstrated that a circumbinary common envelope (CE) is very likely to form shortly after a nova outburst from the interaction of the WD ejecta with the secondary star (MacDonald, 1986; Livio et al. 1990; Kley et al. 1995; MacLeod et al. 2017; Ropke & De Marco 2023). Such an envelope would be a prominent contributor to the spectrum in the epoch of early decline. One of the key characteristics of an envelope is that, like a stellar wind, absorption and emission are linked through radiative transfer processes determined by conditions within the envelope.

Recently, Williams et al. (2022) analyzed the photometry and spectra obtained during decline of the unusual nova V5856 Sgr/2016, which is unique in having remained close to its peak outburst luminosity for more than six years following outburst. Different aspects of its spectra were described in general terms for a common envelope that might explain the unusual behavior of the nova after outburst. However, whereas the models of winds and expanding ejecta that Hauschildt, Shore, Aydi and associates have calculated have been used in the interpretation of novae spectra, little has been done to calculate the expected spectrum of a CE after a nova outburst. We propose here a simplified structure for a CE that could serve as a starting point for treating emission and absorption processes, and that might explain some of the observed spectral characteristics after outburst. We focus on the visible region of the spectrum because that is where the majority of time-resolved data exist.

## 2. Common Envelope Geometry and Emission

Shortly after outburst the visible spectra of most novae are characterized by a continuum with superposed emission lines, some of which have blueward absorption components, i.e., classic P Cygni profiles. The absorption features can be understood as arising from absorption and re-distribution of continuum radiation by an optically thick transition in an expanding gas (Sobolev 1960; Mihalas 1978). If the emission component exceeds the absorption component, creation of photons for that transition, rather than scattering of continuum radiation, is indicated. For a centrally energized, coherent mass of gas the emission lines are formed above the continuum region, where the photosphere radius $r_{ph}$ is defined by where $\tau_{cont} \approx 1$ along the line of sight into the object. This configuration is very similar to that of a normal stellar atmosphere, augmented by the presence of an optically thin 'chromosphere' overlying the continuum and absorption-line formation region.

Absorption lines are produced by an outwardly decreasing temperature gradient with a stellar density distribution determined by hydrostatic equilibrium (Mihalas 1978). Emission lines are produced by an extended, optically thin outer region that can be in the form of a stellar wind that originates in the outer layers of the photosphere, or a detached outer shell as in the case of Be stars. Such is the case for common envelopes, where in eruptive variables the outer layers experience mass loss and are not in hydrostatic equilibrium (Ivanova, Justham, & Ricker 2020). The emission line intensities depend upon the temperature and extent of the overlying optically thin gas, whereas the continuum brightness depends on the surface area and temperature of the photospheric region.

We propose a simple CE model that represents conditions in the postoutburst emitting regions of novae by an initially dense gas that expands. Conceptually, this geometry is similar to the winds treated by Hauschildt et al. (1992, 1994), however its structure corresponds more closely to the CE models that have been computed by Livio et al. (1990), Kley et al. (1995), and MacLeod et al. (2017), in that their density profiles are very different from those dictated by a wind that follows the equation of continuity for a positive velocity gradient. It should be noted that although the MacLeod et al. models represent a luminous red nova mass transfer binary, the primary star is not a degenerate WD.

For our assumed CE geometry, the luminosity of the nova is determined by the Planck function continuum intensity, where $L_{cont} \propto r^2_{ph} T^4_{ph}$, and $r_{ph}$ is the effective photosphere radius. The photosphere effective temperature $T_{ph}$ is set by energetics originating from the inner regions of the gas that produce an outwardly decreasing temperature gradient, thereby accounting for the formation of absorption lines. Simplifying our common envelope geometry in order to define the regions where absorption and emission lines predominate, we consider the effective photosphere region to be overlayed by an optically thin region, i.e., a chromosphere. This allows a basic relation to be established that connects emission line strengths to that of the continuum. Emission from the outer optically thin 'chromosphere' gas surrounding the photosphere consists predominantly of emission lines from ions *i* whose luminosities $L_{em}$ are

$$L_{em} \propto h\nu_i \int n_e n_i \alpha_i^{eff} dV, \qquad (1)$$

where V is the volume of the optically thin gas extending beyond the effective photosphere, $n_e$ and $n_i$ are the electron and ion densities, and $\alpha_i^{eff}$ is the sum of the recombination and collisional excitation coefficient of the emission line (Osterbrock & Ferland 2006).

For the outer optically thin region, rather than postulate its density distribution ad hoc we take the emission-line region to be an extension of the photosphere region by representing its density by a power-law distribution,

$$n(r) = n_{ph} \times (r_{ph}/r)^{\xi}, \qquad (2)$$

where the effective photosphere density $n_{ph}$ and radius $r_{ph}$ are defined by the point where the continuum optical depth into the envelope at 5000 Å, $\tau_{vis}=1$. Since $r_{ph}$ is defined by an integral over distance that involves temperature, ionization, and density, $n_{ph}$ cannot be specified independently of $r_{ph}$. Self-consistent values of $n_{ph}$ and $r_{ph}$ need to be found iteratively, and may be done straightforwardly via the procedure described in Appendix A. The condition that we invoke for common envelopes, viz., that the density of the optically thin outer region of gas is related to the inner optically thick gas producing the continuum, i.e., not isolated from it, is for us a defining feature for a nova CE. It defines the general structure of the CE, and previous models of CE's associated with mass transfer binaries have invoked this condition (Ivanova, Justham, & Ricker 2020; Ropke & De Marco 2023).

The effective photosphere region is defined where $\tau_{vis}=1$ into the ejecta, with

$$\tau_{vis} = \int_{r_{ph}}^{\infty} n(r) \Sigma x_i \, a_i \, dr , \qquad (3)$$

where $x_i$ is the fractional abundance of ion *i* relative to the total density, and $a_i$ is the absorption cross section at 5000 Å. The spectra of novae in early decline normally exhibit low ionization species, e.g., $Na^o$, $O^o$, and $Fe^+$, even when higher ionization and excitation transitions of $H^o$ and $He^o$ also appear. In the latter case, P Cygni profiles of the higher excitation species can show different radial velocity absorption features than the lower ionization species. Detailed calculations of ionization and temperature are necessary to determine the sources of optical depth into the CE. For the relatively dense region of the gas where the visible continuum and absorption lines are formed, we assume initially that local thermodynamic equilibrium (LTE) conditions apply. For densities taken from the calculations of novae CE models by Livio et al. (1990) and Kley et al. (1995), the continuous absorption coefficients expected to predominate in the visible are most likely $H^-$ bound-free, $H^o$ Balmer continuum, and $H^+$ free-free absorption, as has been found for the accretion disks and ejecta of cataclysmic variables with similar physical conditions (Williams 1980, Fig. 4; Metzger et al. 2021).

The $H^-$ cross section in the visible is $a(H^-) \sim 3 \times 10^{-17}$ cm$^2$ (Wishart 1979), and the cross sections for both $H^o$ Balmer and $H^+$ free-free absorption are of order $6 \times 10^{-18}$ cm$^2$ (Mihalas 1978). Much less certain are the relative abundances of the particles involved because the absorption coefficients depend upon the presence of free electrons, and these likely come from low ionization level metals. Analyses of novae continua in early decline normally indicate photospheric temperatures around 8,000 K (Bath & Harkness 1989), although this temperature is not necessarily applicable to the optically thin emission region. The excitation of $H^o$ to the n=2 level will produce $H(n=2)/H \leq 10^{-3}$ unless the temperature exceeds $10^4$ K. The consistent strengths of O I $\lambda\lambda$7773 and 8446 in early decline suggests relatively low ionization of H and He, so the free electrons accessible to form $H^-$ are likely due primarily to $Fe^+$, the most abundant of the heavy elements with low ionization potentials. Depending on the Fe/H abundance, an approximate value of $H^-/H \leq 10^{-3}$ may be typical. Thus, allowing for uncertainties in the temperature and ionization, whose values do require construction of detailed models, we adopt the combined fractional abundance of absorbers to be $\sim 10^{-3}$ of the total density, and the combined absorption cross sections in the visible to be $<a_i>=10^{-17}$ cm$^2$.

Using the density distribution from eqn. (2) and the approximate parameter values above, the $\tau_{vis}=1$ requirement constrains the density-radius relationship for the effective photosphere to be,

$$n_{ph} \, r_{ph} \approx 10^{20} \, (\xi - 1). \qquad (4)$$

For a typical nova the Livio et al. and Kley et al. models show the densities of common envelopes to be highest by more than an order of magnitude within the radius of the secondary star photosphere. As the secondary sweeps around its orbit it therefore creates an effective photosphere for the CE at its orbital radius, i.e., with $r_{ph} \approx 3 \times 10^{10}$ cm for a low mass main sequence star with an orbital period of some hours. Treating the CE as a single coherent entity rather than a collection of discrete clumps, this leads to a photosphere density of $n_{ph} \approx 10^{10}$ cm$^{-3}$ for low values of $\xi<5$. This density is too low to justify the assumption of LTE, and is slightly lower than that necessary to explain the frequently observed O I $\lambda$7773 emission line, whose excitation via H I Ly-$\beta$ resonance fluorescence of O I quintet levels requires a density of at least $10^{10.5}$ cm$^{-3}$ (Kastner & Bhatia 1995), suggesting that higher values of $\xi$ are appropriate. In their model atmosphere calculations with a wind, Hauschildt et al. (1994) came to this same conclusion, i.e., a steep density gradient with $\xi \approx 15$ was necessary for the outer optically thin gas in order to produce an acceptable fit to the nova V1974 Cygni/1992 early spectra. If a common envelope forms shortly after outburst, steep density gradients are required for the outer region of the CE in order to allow sufficiently high densities for continuum and absorption line formation with emission lines also prominent.

If the density of gas in the general region of the effective photosphere can be represented adequately by the eqn (2) power-law distribution, a straightforward relationship exists between the equivalent width of an emission line and the radius and density of the photosphere. Taking the Balmer H$\beta$ line as an example, the equivalent width of the line, i.e., the intensity of the line compared with the continuum intensity, is defined as EW$_{H\beta}$ = L$_{H\beta}$/(L$_{cont}$ $\Delta\nu_1$), where $\Delta\nu_1$ is the frequency width of a 1 Å interval at H$\beta$. The continuum luminosity is approximately L$_{cont}$ = $4\pi^2 r_{ph}^2 B_\nu(T_{ph})$, where B$_\nu$ is the Planck function, although it should be noted that at densities below ~$10^{13-15}$ cm$^{-3}$, the assumption of LTE is likely to break down and the continuum intensity will deviate from the Planck function.

Using eqn. (1) for the luminosity of H$\beta$ in terms of its effective recombination coefficient $\alpha_{H\beta}^{eff}$ (Osterbrock & Ferland 2006) and the eqn. (2) power-law density distribution, the equivalent width can be expressed in terms of the temperature, radiation field, abundances, and ionization level of the gas, as

$$\text{EW}_{H\beta} = h\nu_\beta\, x_H^2\, \alpha_\beta^{eff}/[(2\xi - 3)\, \pi\, B_\nu(T_{ph})\, \Delta\nu_1] \times n_{ph}^2\, r_{ph}, \qquad (5)$$

where $x_H$ is the fraction of the density that is ionized hydrogen, H$^+$. The $n_{ph}$ and $r_{ph}$ are self-consistent values of the photosphere density and radius that satisfy the requirement that $\tau_{vis}$=1. For a power-law density distribution the integrand for optical depth in eqn. (3) is most dependent on the region having highest density, which is close to the photosphere. This results in a power-law density distribution strongly favoring emission lines being strongest near the photosphere continuum----a feature that may explain how early decline spectra can change rapidly. The determination of emission line equivalent widths serves as an important test of the validity of a common envelope model.

### 3. Time-Resolved Spectral Features

Models based on different geometries have been proposed for novae, including discrete non-interacting clumps, collimated jets, and colliding shells. It is clear that separate components of ejecta do collide with each other in decline, producing shocks that power postoutburst activity (Steinberg & Metzger 2020). Spectra obtained when novae are in the Fe II spectroscopic phase frequently show lines having prominent, multiple absorption components with different radial velocities (Izzo et al. 2015; Aydi et al. 2020a). Multiple components having different velocities may be the result of episodic mass loss events from a common envelope as it dissipates into distinct ejecta components.

We have examined a collection of time-resolved novae spectra to look for characteristics that might be signatures of ejecta geometry. Over the past several decades two extensive databases of postoutburst novae spectra have been posted online. The first is F. Walter's SMARTS Atlas of Southern Novae[1] (Walter et al. 2012). The second is the world-wide volunteer Astronomical Ring for Amateur Spectroscopy[2] (ARAS) group (Teyssier 2019). Using different telescopes and instruments, they have amassed a useful collection of spectra that are publicly available. Walter and colleagues have used different telescopes and spectrographs on Cerros Tololo and Pachon in Chile. ARAS spectra have been obtained by many observers using different instruments and spectral resolutions and in varying photometric conditions. A unique value of the ARAS database is that with its numerous observers situated globally, there can be more than ten separate spectra of a nova obtained in a single night.

Of the many characteristics displayed by novae spectra after outburst several may be supportive of their formation in a CE. The first is the presence in early spectra of Balmer and Fe II lines that have prominent P Cygni profiles, with both emission and absorption components having a range of strengths and widths. A substantial literature exists on the formation of P Cygni line profiles for different geometries and conditions (Rottenberg 1952; Lucy 1971; Klein & Castor 1978; Castor & Lamers 1979; & Hillier 1991). The calculations show such profiles to be a consequence of an optically thick expanding medium, with the profile shapes determined primarily by the assumed density and velocity laws with distance. In comparing the high-resolution profiles observed by Aydi et al. (2020a), displayed in their Figs. 1 and 2, with the P Cygni profiles computed in the above cited studies, all have shapes that can be accounted for by an expanding common envelope. Multiple absorption components are created by density peaks with different velocities, which may be indicative of multiple ejection or mass loss episodes.

---

[1] https://www.astro.sunysb.edu/fwalter/SMARTS/NovaAtlas/atlas.html
[2] https://aras-database.github.io/database/novae.html

In particular, for a nova CE after outburst that is not in hydrostatic equilibrium and likely to be losing mass, one does expect strong absorption to occur at velocities in the range of the CE escape velocity, which for a radius of 1 $R_\odot$ with a 1 $M_\odot$ primary WD should be in the range 4-6×$10^2$ km/s (Kley et al. 1995). Of the dozen novae displayed by Aydi et al., half of them do have absorption in this velocity range at the epochs displayed. Absorption in the other half occurs at higher velocities, which may mean that postoutburst activity on the WD has filled the CE with high-energy radiation and ejecta, causing sudden high-velocity mass loss from the CE.

A second notable feature of postoutburst spectra that could be a favorable indicator of novae CE's is that fundamental changes in the basic spectra of novae occasionally occur quite rapidly, often in only a day or two. These changes take place generally in the period of early decline, primarily when the brightness of the nova is still within a few magnitudes of its peak visible luminosity. In these instances, the spectrum typically converts from predominantly emission lines to absorption lines that strengthen in a matter of a few days, and with the exception of H$\alpha$, the emission lines fade into the continuum. Often the spectrum then reverts back to a similar emission-line configuration after an interval of a week or more. Rapid spectrum changes, which for convenience we will call *shifts*, contrast with the normal evolution of novae spectra after forbidden lines first appear, when change proceeds gradually via a regular evolution in the relative intensities of emission lines.

During rapid shifts some novae change visible brightness, but there is no systematic trend evident in either the direction or extent of brightness variations during shifts for the majority of objects. In Figures 1-5 we show examples of novae whose spectra experienced relatively rapid shifts at a time the brightness was near peak visible luminosity. The spectra displayed are reproductions of FITS files posted on the public ARAS website, where information is also provided for each observation on the day it was obtained. The range in spectral resolution used for some of the displayed spectra may cause detection of some of the weakest, narrowest spectral features to be compromised in some spectra. The light curves of the five novae are displayed in Fig. 6, with data taken from the AAVSO website[3]. Vertical marks denote the dates when the spectra shown in the figures were taken.

There are some common features that novae exhibit related to shifts. Initially, the prominent features in the spectra are Balmer and Fe II multiplet 42 emission lines that often have P Cygni profiles with emission components dominant. That is, shifts normally occur when novae are in the 'Fe II' spectral phase (Williams 2012; Aydi et al. 2024). Within 1-3 days absorption lines appear and become stronger as the emission components diminish in strength, often leaving H$\alpha$ as the only prominent emission feature.

---

[3] https://www.aavso.org/LCGv2/

Exceptions do occur, e.g., the Balmer and Fe II lines for V1405 Cas retained their P Cygni profiles with strong absorption components before, during, and after the entire shift. Also, Nova V613 Sct deviates from the procedure described above in that the sudden transformation of its spectrum from predominant absorption lines to emission lines that occurred within 24 hours on 1 July was not preceded by an emission spectrum. This may be due to the fact that the nova was discovered after peak brightness. The 30 June spectrum shown in Fig. 3 was taken just 30 hours after the nova's discovery, and the already declining light curve suggests that the nova was likely emitting an Fe II emission spectrum before its discovery.

During spectral shifts the absorption lines tend to be the same transitions found in transient heavy element absorption (*thea*) systems (Williams et al. 2008), although the radial velocities of *thea* systems often differ from that of the strongest absorption component of lines having P Cygni profiles. Spectral shifts may be the origin of *thea* systems in most circumstances. The strong, dominant absorption components of Balmer and Fe II P Cygni profiles represent absorption by the expanding, dispersing common envelope. *Thea* systems consisting of transitions of Sr, Ba, Y, Ti and other Fe-peak elements may represent ejection events from the CE driven by episodic bursts of $\gamma$-rays, strengthening their direct association with the secondary star.

The main phase for each *shift* is the interval where absorption lines are the dominant feature of the spectrum, which typically last of order 1-2 weeks. The spectrum then changes over an interval of a few days, back to a predominant emission spectrum that is similar to that which prevailed before the shift occurred. The most interesting aspect of spectrum shifts remains the suddenness with which the visible spectra of novae change their basic structure.

## 4. Gamma Ray Emission and Spectrum Changes

The cause of sudden spectrum shifts is clearly due to major changes in the emitting region conditions. One possible source of energy that could initiate shifts is the emission of high energy $\gamma$-rays that have been detected in postoutburst novae. Since its launch in 2008, the Large Area Telescope (LAT) on *Fermi* Gamma-ray Telescope has conducted a sky survey to identify sources that emit gamma rays. Of the roughly 165 Galactic novae that have been confirmed spectroscopically since launch, approximately 20 novae have been identified as sources of $\gamma$-ray emission in the list maintained by K. Mukai[4]. Characteristics of the $\gamma$-ray novae have been discussed by Franckowiak et al. (2018) and Chomiuk, Metzger, & Shen (2021), with the striking fact that most novae detections have been made just above the LAT detection limit. Detections have often required integration on the object field of view over several days in order to achieve sufficient signal-to-noise to be designated as statistically valid sources. The LAT detection limit does discriminate against the detection of $\gamma$-rays from the more distant novae.

---

[4] https://asd.gsfc.nasa.gov/Koji.Mukai/novae/novae.html

Observed fluxes over the 20 MeV-300 GeV energy range of LAT for all detections represent a small fraction, <1%, of the radiant energy of the novae, which generally amounts to roughly the Eddington luminosity of a 1 $M_\odot$ object. The γ-ray luminosities of the detected novae show a wide range of values, such that a majority of novae could generate γ-rays following outburst at luminosities up to 1% the luminosity of the nova, yet still remain undetected with current facilities (Franckowiak et al. 2018).

A key correlation has been found between the γ-ray intensity and optical brightness of novae (Li et al. 2017; Aydi et al. 2020b). For novae for which γ-ray flux has been detected, the positive correlation with visible brightness has been suggested to be due to both γ-rays and visible radiation being produced by interactions within the ejecta, so that "the majority of the optical light comes from reprocessed emission from shocks rather than the white dwarf" (Li et al. 2017). The inevitable collision between ejecta components that have different velocities makes this hypothesis highly likely (Metzger & Pejcha 2017; Aydi et al. 2020a).

In addition to the above calculations that support the feasibility of ejecta shocks producing observed gamma rays, a correlation between spectrum changes and gamma ray intensity could be indicative of a more centralized source for the gamma rays. We suggest that consideration be given to the fact that postoutburst γ-rays might originate from activity on or near the white dwarf. There has been no expectation that novae white dwarfs produce γ-rays, but the same could be said of the sun, where they have been observed at energies exceeding 100 Gev. Linden et al. (2022) recently completed a survey of Fermi LAT solar observations over a complete 11 yr solar cycle, and found an anti-correlation between solar activity and γ-ray emission that was interpreted in terms of solar magnetic fields producing the observed fluxes. Further analysis of the data by Arsioli & Orlando (2024) revealed an asymmetry in solar disk emission that further linked observed γ-ray emission to solar magnetic fields, rather than the re-processing of galactic high energy cosmic rays. Banik et al. (2023) have proposed that acoustic-shock disturbances move upward in the solar atmosphere, producing the γ-rays. Such a process could take place above the surface of the nova WD.

The magnetic fields of novae WDs are expected to be many orders of magnitude greater than that of the sun. In polars like AM Her, and novae as V1500 Cygni/1975 and DQ Her/1934, observations have demonstrated field strengths in excess of $10^7$ gauss (Schmidt, Stockman, & Grandi 1983; Schmidt & Stockman 1991). Thus, centrally localized γ-ray emission associated with activity on novae WDs does have credibility. It could ionize and heat the gas in the outer layers of the common envelope. Such a situation would also produce a positive correlation between γ-ray and visible brightness. When incident upon the ejecta, whether in the form of a CE or discrete shells or globules, γ-rays could definitely generate rapid spectroscopic changes because the small cross section of matter to γ-rays allows them to reach every region throughout a CE and produce residual ionization that can dictate both the position of and conditions in the effective photosphere and overlying emission-line region.

As a test of this hypothesis, we have checked for possible $\gamma$-ray detections the novae shown in Figs. 1-5 that experienced rapid spectrum shifts:

1. *V5855 Sgr/2016c*: After discovery on 20 October 2016, $\gamma$-rays were detected between 28 October – 1 November 2016 (Li & Chomiuk 2016). It is clear from Fig. 1 that a major spectrum *shift* occurred early in decline, between 22-26 October 2016, shortly before the $\gamma$-rays were detected. By 23 October the spectrum was entirely in absorption, with the exception of an H$\alpha$ emission component. The *shift* concluded with a very rapid reversal in 24 hours between 25-26 October, in which its absorption spectrum returned to being dominated by emission lines. The $\gamma$-ray detection occurred within 2 days of the spectral shift. Therefore, the two events are possibly related, although not quite coincident with each other.
2. *V5856 Sgr/2016d*: The spectral evolution of this nova was discussed in detail by Williams et al. (2022). As shown in Fig. 2, ARAS spectra show a spectrum shift in the interval 29 October – 18 November 2016. The initial spectra display emission with only a few absorption features present. Between 30 October – 9 November H$\alpha$ is the only emission feature. By 18 November a prominent emission spectrum has emerged. Li et al. (2016) detected $\gamma$-rays with Fermi LAT from 8-17 November, during the time the spectrum was transitioning back to the emission spectrum. So, the spectral shift and $\gamma$-ray emission for this nova did overlap in time.
3. *V613 Sct/2018*: Dominated by absorption lines on its 29 June 2018 discovery, the spectrum of this nova would be unusual if it were at peak brightness. But, because no pre-discovery observations of its immediate area of the sky were reported within weeks of its discovery, it was likely first observed after maximum luminosity, at a time when emission lines were probably prominent. The ARAS spectra in Fig. 3 show the spectrum to consist predominantly of absorption lines until early 1 July. However, later that same day within 20 hours, the absorption largely disappeared, abruptly replaced by strengthening emission lines. The spectrum retained this appearance with minimal absorption until observations ended in late July. No $\gamma$-rays were detected from this object, so there is no direct evidence that they played a role in the rapid change of the visible spectrum. However, because this nova was visibly the faintest of the group, $\gamma$-rays may well have been present at any time with normal luminosity, but remained undetected by the Fermi LAT.
4. *V1674 Her/2021*: Discovered on 12 June 2021, this relatively bright nova was immediately observed by Fermi LAT and by ARAS spectroscopists. Li (2021a, 2021b) reported a relatively strong $\gamma$-ray flux over 12-13 June, however no further detections were reported, indicating that the $\gamma$-rays weakened quickly. The ARAS spectra shown in Fig. 4 show the nova to be in the He/N phase at discovery with rather broad absorption features indicating high velocities for the emitting gas, with Fe II emission also present. The rapid drop in the visible light curve shown in Fig. 6 indicates the nova might not have been discovered at peak brightness. During 13 June the spectrum began a rapid change from having strong P Cygni profiles to a dominant, broad emission spectrum. This happened at the same time as the brief, strong γ-ray emission, so there is good reason to believe the γ-rays and major spectral change were causally related.

5. *V1405 Cas/2021*: The correlation between high-energy γ-ray and visible luminosity noted by Li et al. (2017) is quite pronounced for this nova, discovered in outburst on 18 March 2021. It was not detected as a γ-ray source until 21-24 May 2021 (Buson et al. 2021), at peak visible brightness more than 60 days following outburst, as shown in Fig. 6. The nova experienced a change in spectrum between approximately 8-16 May that although not as pronounced as the shifts observed in other novae, does show absorption lines becoming prominent for more than one week before fading significantly. The detection of γ-rays, the relatively rapid rise to maximum visible brightness beginning around 3 May, and the appearance and then disappearance of absorption lines, especially in the region below 4700 Å, together in May after such an abnormally long delay time following outburst, does suggest a possible correlation between the three events.

These data represent evidence that γ-ray emission may be correlated not only with visible brightness of novae, but also with the rapid changes that are observed in their spectra.

## 5. The Summing Up

A common envelope is an expected formation for novae soon after outburst when outburst radiation and ejecta impact the secondary star. Because imaging is not yet capable of resolving the detailed structure of mass transfer binaries, spectral information is the best means of demonstrating the presence of a CE in the early decline period. We propose a simple process to model the spectra from a CE for an assumed power-law density distribution. The observed presence of strong emission lines together with absorption lines requires a step density distribution for an optically thick CE. Models with different density and velocity laws can be constructed by an iterative process, outlined in Appendix A, that allows both emission and absorption line strengths to be determined, that can be compared with observations.

Time resolved spectroscopic data show that the spectra of novae in their early decline period often show sudden changes in their basic nature. Many of these *shifts* begin with predominant emission lines that change in a matter of a few days to an absorption spectrum, and then revert back to an emission-line spectrum after a period of days-to-weeks. This behavior may be accounted for straightforwardly by a common envelope, especially since time-resolved spectra observe spectral shifts to occur at or near times they may be activated by γ-ray emission.

The suggestion that shocks from collisions between ejecta components produce high energy gamma radiation and conditions that explain key aspects of novae spectra is surely valid. That does not preclude other sources of high energy from also being important, such as activity on the primary WD. The relatively quiescent sun, with magnetic fields orders of magnitude less than those of WDs in novae, produces a flux of very high energy $\gamma$-rays from solar flares. It would hardly be surprising if postoutburst activity on WD's were able to provide for the $\gamma$-ray luminosities observed for novae which, like that of the sun, are clearly variable.

The authors are indebted to Dr. Fred Walter and the world-wide group of ARAS observers for their diligent acquisition of spectra that they reduce and post online for interested researchers. We thank Dr. K. Mukai for maintaining his very useful list of galactic novae that allows samples of novae with similar characteristics to be identified straightforwardly. We also acknowledge the important work of the American Association of Variable Star Observers, who have provided the photometry for the light curves displayed in Fig. 6.

## Appendix A

   A simplified model can be constructed for a spherical body of gas that represents the continuum and emission line characteristics of a nova common envelope. An iterative procedure is invoked where the gas is presumed to be optically thick in the continuum, and in local thermodynamic equilibrium (LTE). It begins by arbitrarily specifying an initial radius for the CE photosphere, $r_{ph}$, an initial photosphere density, $n_{ph}$, and an assumed temperature distribution, $T(r)$. The density distribution out from the photosphere is prescribed in terms of $n_{ph}$ by eqn. (2), with the power-law variable $\xi$ initially selected.

   The optical depth $\tau_{vis}$ at 5000 Å is computed with the above parameters along the line of sight from infinity into the center of the gas until the point where $\tau_{vis}=1$ is reached. This requires assumed element abundances that enable the continuum absorption coefficient to be determined from ionization and excitation calculations using well known LTE relationships. The values of density $n_1$ and CE effective photosphere radius $r_1$ at this point will not agree with the initially specified values of $n_{ph}$ and $r_{ph}$.

   The goal of the procedure is to iterate the optical depth calculation by adjusting the initial parameter values so subsequent models converge to self-consistent values of the photosphere density and radius, and *with parameter values that produce the emission line equivalent widths that are observed for novae*. Determination of the temperature distribution from heating and cooling processes is important. The calculation of optical depth is sufficiently straightforward that many iterations should be able to be performed on a computer very rapidly, allowing convergence to be achieved.

   The key is focusing on the parameter(s) that are most sensitive to determining the emission line strengths relative to the continuum. For a slowly varying, flat density distribution the optical depth builds up from the low outer densities to the point where $\tau_{vis}=1$ before densities $n \geq 10^{15}$ cm$^{-3}$ are reached that justify the assumption of LTE. This problem should be avoided by setting a lower outer limit to the assumed photosphere density, $n_{ph}$. This causes the power-law density parameter $\xi$ to be a key parameter to focus on when building a suite of useful common envelope models. Also, there is no reason that the eqn (2) power-law density must be adhered to. The power-law distribution can be modified by an alternative distribution.

# References

Arsioli, B. & Orlando, E. 2024, ApJ, 962, 52

Aydi, E., Chomiuk, L., Izzo, L. et al. 2020a, ApJ, 905, 62

Aydi, E., Chomiuk, L., Strader, J., Sokolovsky, K.V., Williams, R.E., et al. 2024, MNRAS, 527, 9303

Aydi, E., Sokolovsky, K.V., Chomiuk, L., Steinberg, E., Li, K-L., et al. 2020b, Nat Astron, 4, 776

Bath, G.T. & Harkness, R.P. 1989, in *Classical Novae*, ed. M.F. Bode & A. Evans (New York: J. Wiley & Sons), p. 61

Beals, C.S. 1931, MNRAS, 91, 966

Buson, S., Cheung, C.C., Jean, P., & Fermi LAT Collaboration. 2021, ATel #14658

Castor, J.I. 1972, ApJ, 178, 779

Castor, J.I. & Lamers, H.J.G.L.M. 1979, ApJS, 39, 481

Chomiuk, L., Metzger, B. D., & Shen, K. J. 2021, ARAA, 59, 391

Franckowiak, A., Jean, P., Wood, M., Cheung, C.C., & Buson, S. 2018, A&A, 609, 120

Hachisu, I. & Kato, M. 2022, ApJ, 939, 1

Hauschildt, P.H., Starrfield, S., Austin, S. et al. 1994, ApJ, 422, 831

Hauschildt, P.H., Starrfield, S., Shore, S.N. et al. 1995, ApJ, 447, 829

Hauschildt, P.H., Wehrse, R., Starrfield, S., & Shaviv, G. 1992, ApJ, 393, 307

Hillier, D.J. 1991, A&A, 247, 455

Ivanova, N., Justham, S., & Ricker, P. 2020, *Common Envelope Evolution* (Bristol: Inst. of Phys)

Izzo, L., Della Valle, M., Mason, E. et al. 2015, ApJL, 808, L14

Kastner, S.O. & Bhatia, A.K. 1995, ApJ, 439, 346

Klein, R.I. & Castor, J.I. 1978, ApJ, 220, 902

Kley, W., Shankar, A., & Burkert, A. 1995, A&A, 297, 739

Li, K-L. 2021a, ATel #14705

Li, K-L. 2021b, ATel #14707

Li, K-L. & Chomiuk, L. 2016, ATel #9699

Li, K-L., Chomiuk, L., Strader, J. et al. 2016, ATel #9771

Li, K-L., Metzger, B.D., Chomiuk, L., Vurm, I., Strader, J., et al. 2017, Nat Astron, 1, 697

Linden, T., Beacom, J.F., Annika, H.G.P., Buckman, B.J., Zhou, B. et al. 2022, PhysRevD, 105, 063013

Livio, M., Shankar, A., Burkert, A., & Truran, J.W. 1990, ApJ, 356, 250

Lucy, L.B. 1971, ApJ, 163, 95

MacDonald, J. 1986, ApJ, 305, 251

MacLeod, M., Macias, P., Ramirez-Ruiz, E. et al. 2017, ApJ, 835, 282

McLaughlin, D.B. 1942, ApJ, 95, 428

Metzger, B.D. & Pejcha, O. 2017, MNRAS, 471, 3200

Metzger, B.D., Zenati, Y., Chomiuk, L. et al. 2021, ApJ, 923, 100
Mihalas, D. 1978, Stellar Atmospheres (W.H. Freeman: San Francisco), p. 310
Osterbrock, D.E. & Ferland, G.J. 2006, *Astrophysics of Gaseous Nebulae and Active Galactic Nuclei*, 2nd ed. (Sausalito: University Sci. Books)
Payne-Gaposchkin, C. 1957, *The Galactic Novae* (Amsterdam: North-Holland Publ. Co.)
Ropke, F.K. & De Marco, O. 2023, LRCA, 9, 2
Rottenberg, J.A. 1952, MNRAS, 112, 125
Schmidt, G.D. & Stockman, H.S. 1991, ApJ, 371, 749
Schmidt, G.D., Stockman, H.S., & Grandi, S.A. 1983, ApJ, 271, 735
Shore, S.N. 2012, BASI, 40, 185
Shore, S. N., Mason, E., Schwarz, G.J. et al. 2016, A&A, 590, 123
Sobolev, V.V. 1960, *Moving Envelopes of Stars* (Cambridge: Harvard Univ. Press)
Sobolev, V.V. 1969, 'A theoretical study of galactic novae', VisAstr, 11, 181
Steinberg, E. & Metzger, B.D. 2020, MNRAS, 491, 4232
Teyssier, F. 2019, CoSka, 49, 217
Walter, F.M., Battisti, A., & Towers, S.E. et al. 2012, PASP, 124, 1057
Williams, R. 2012, AJ, 144, 98
Williams, R.E. 1980, ApJ, 235, 939
Williams, R., Mason, E., Della Valle, M., & Ederoclite, A. 2008, ApJ, 685, 451
Williams, R., Walter, F.M., Rudy, R.J., Munari, U., Luckas, P. et al. 2022, ApJ, 941, 138
Wishart, A. W. 1979, MNRAS, 187, 59

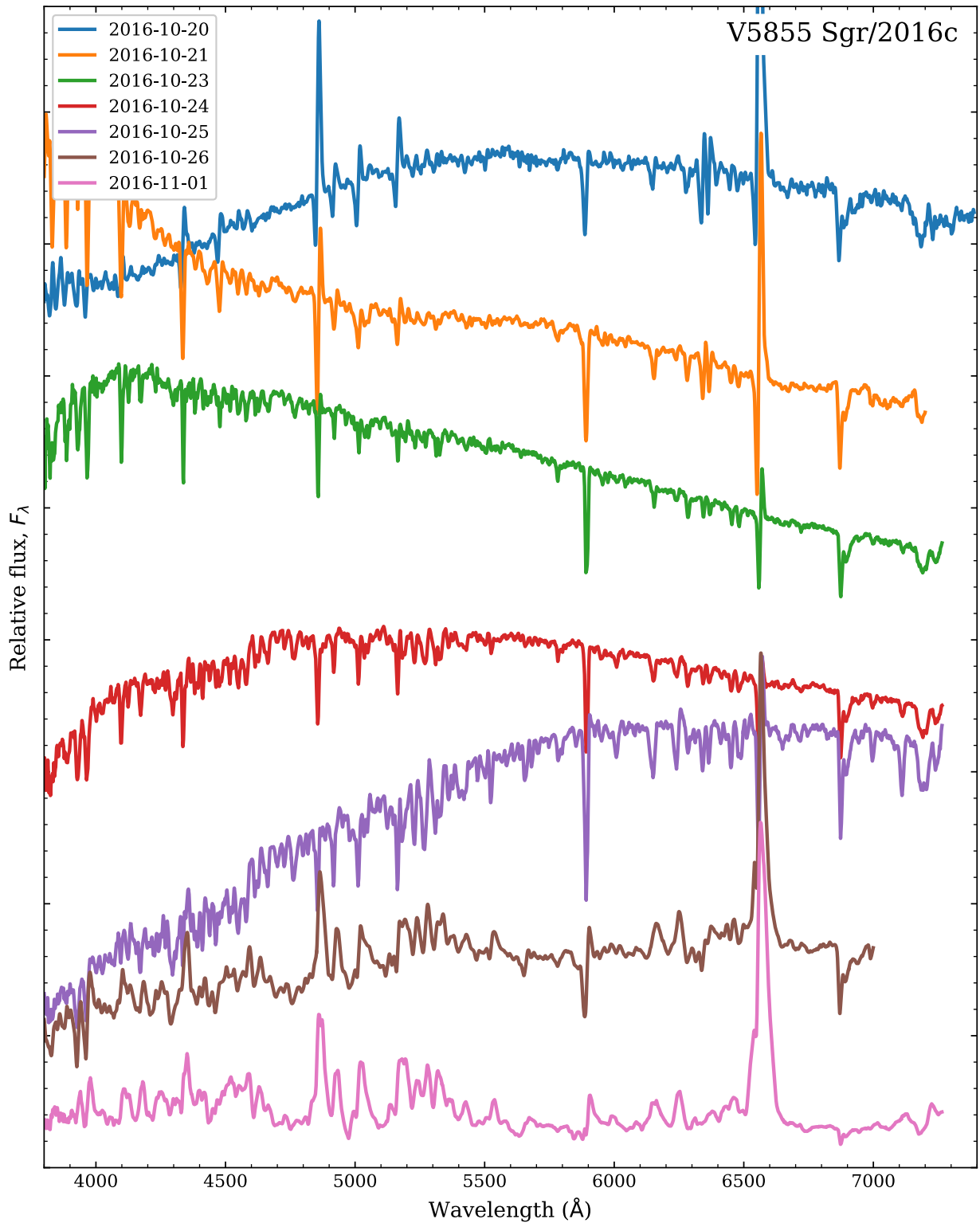

Figure 1 – ARAS spectra of V5855 Sgr shortly after outburst. Except for H$\alpha$, the P Cygni profiles present in the initial 20 October spectrum have disappeared by 23 October. The spectrum suddenly emerging on 26 October shows emission lines prominent, with the large majority of absorption lines present on 25 October having disappeared within 24 hours.

Figure 2

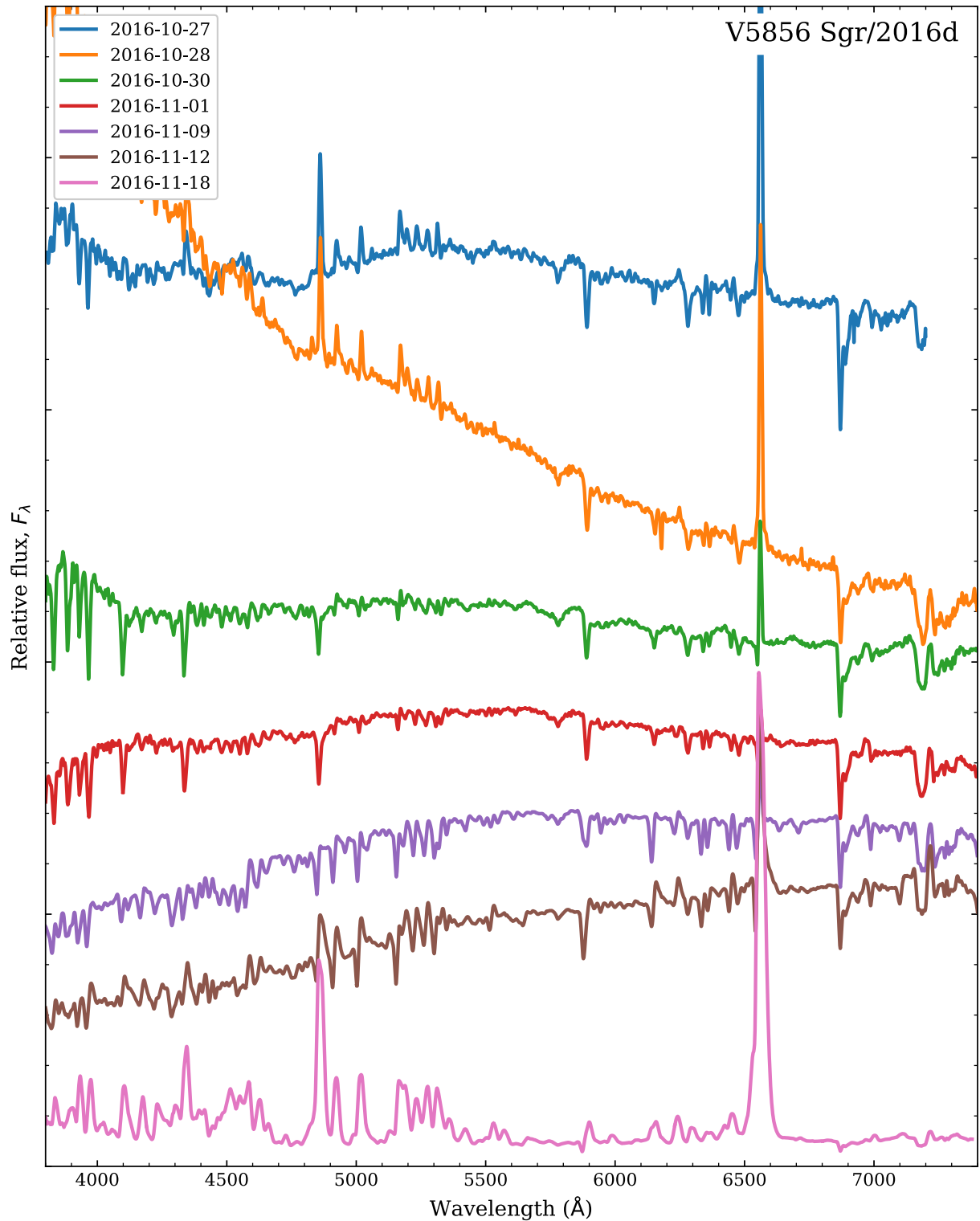

Figure 2 – After an initial Fe II-phase emission spectrum, between 28-30 October the V5856 Sgr spectrum changed from emission to a dominant absorption spectrum. The absorption lines strengthened and persisted until 12 November, when emission components re-appeared in the form of P Cygni profiles. By 18 November the spectrum had reverted to its initial Fe II emission structure.

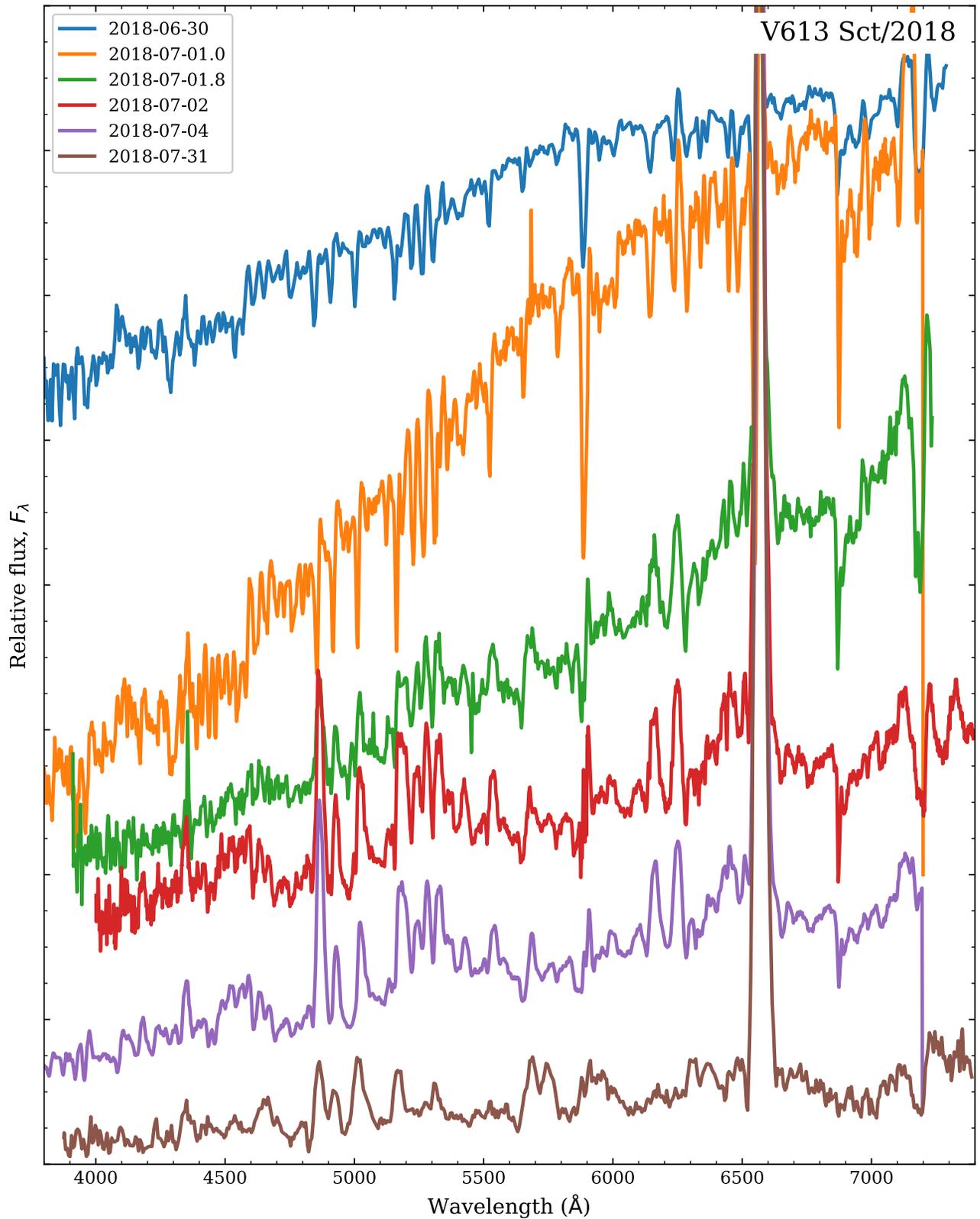

Figure 3 – ARAS spectra for V613 Sct show this nova's spectrum to have transformed significantly within a 20-hour period on 1 July from very prominent, narrow absorption lines to an emission spectrum. During the 1 July transformation interval, the visible brightness may have decreased of order 0.5 mag, as shown in Fig. 6.

Figure 4

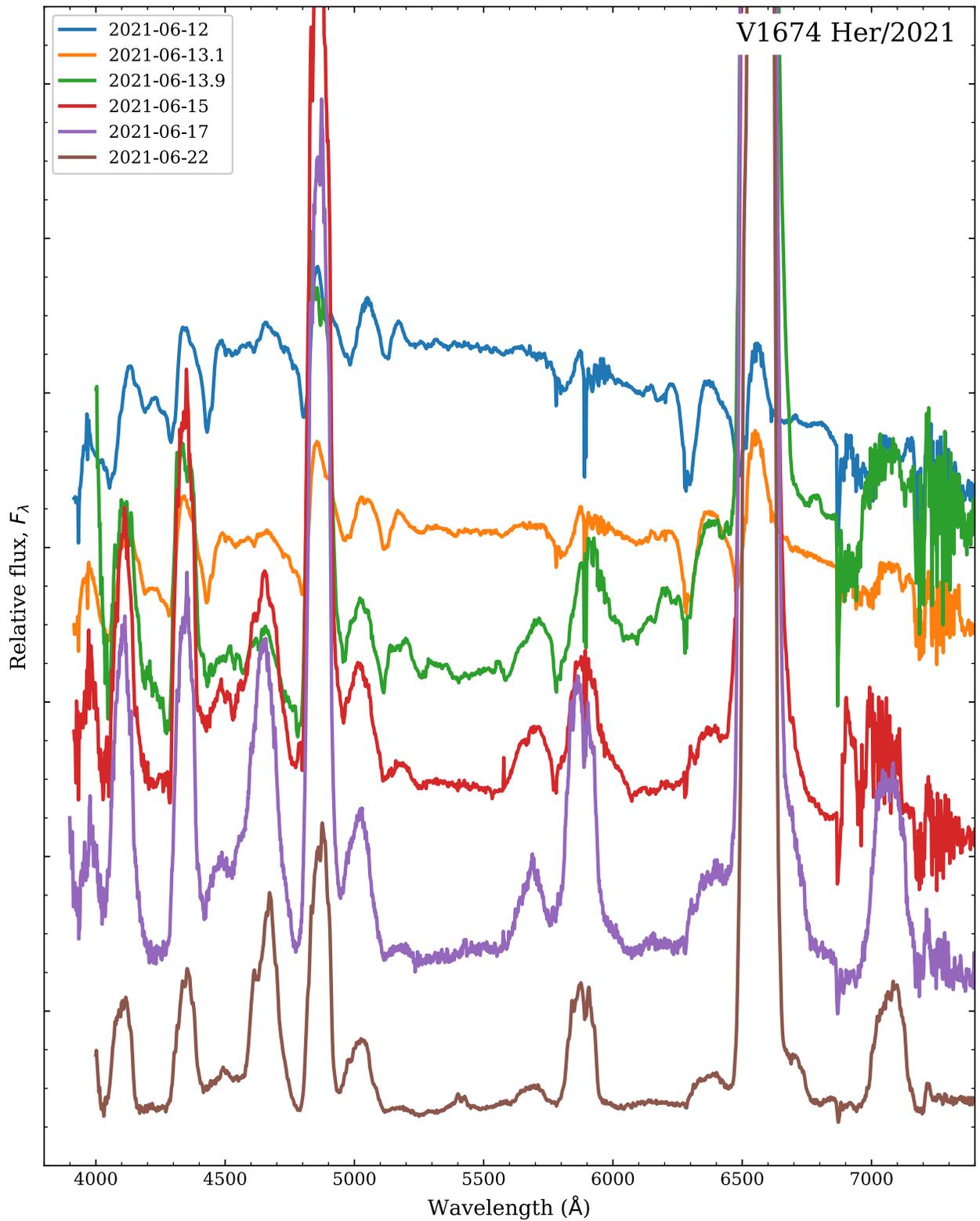

Figure 4 – V1674 Her displayed a He/N-phase spectrum during its rapid decline in brightness. Strong, broad absorption lines initially present in these ARAS spectra disappeared rapidly on 13 June as the spectrum became dominated by strong emission features.

Figure 5

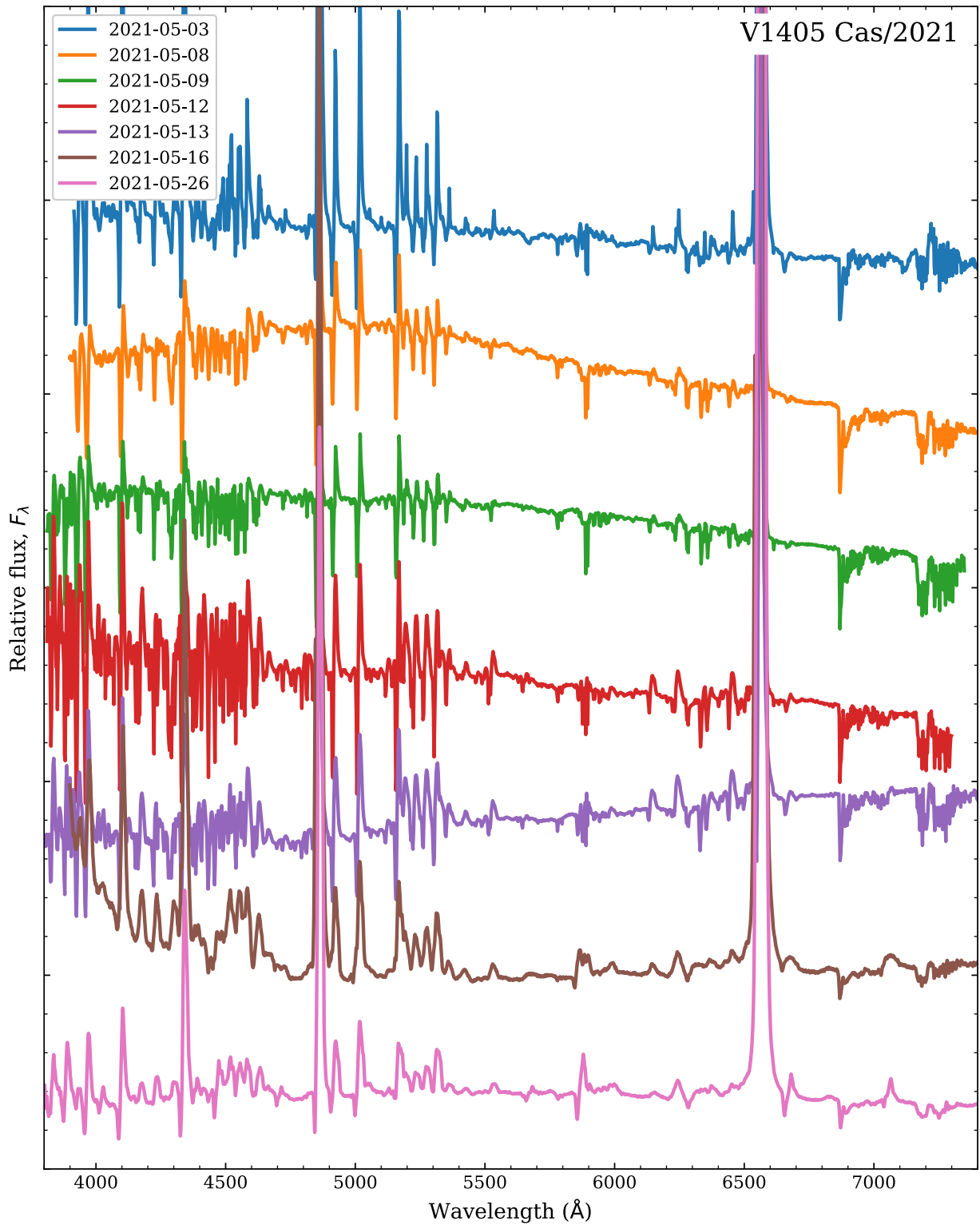

Figure 5 – Nova V1405 Cas displayed classic Fe II-phase spectra in these ARAS spectra with narrow Fe II P Cygni profiles having prominent absorption components until their disappearance on 16 May. Non-Balmer emission components weakened significantly between 3-8 May, as narrow absorption features appeared and strengthened, especially below 4700 Å, until their disappearance.

Figure 6

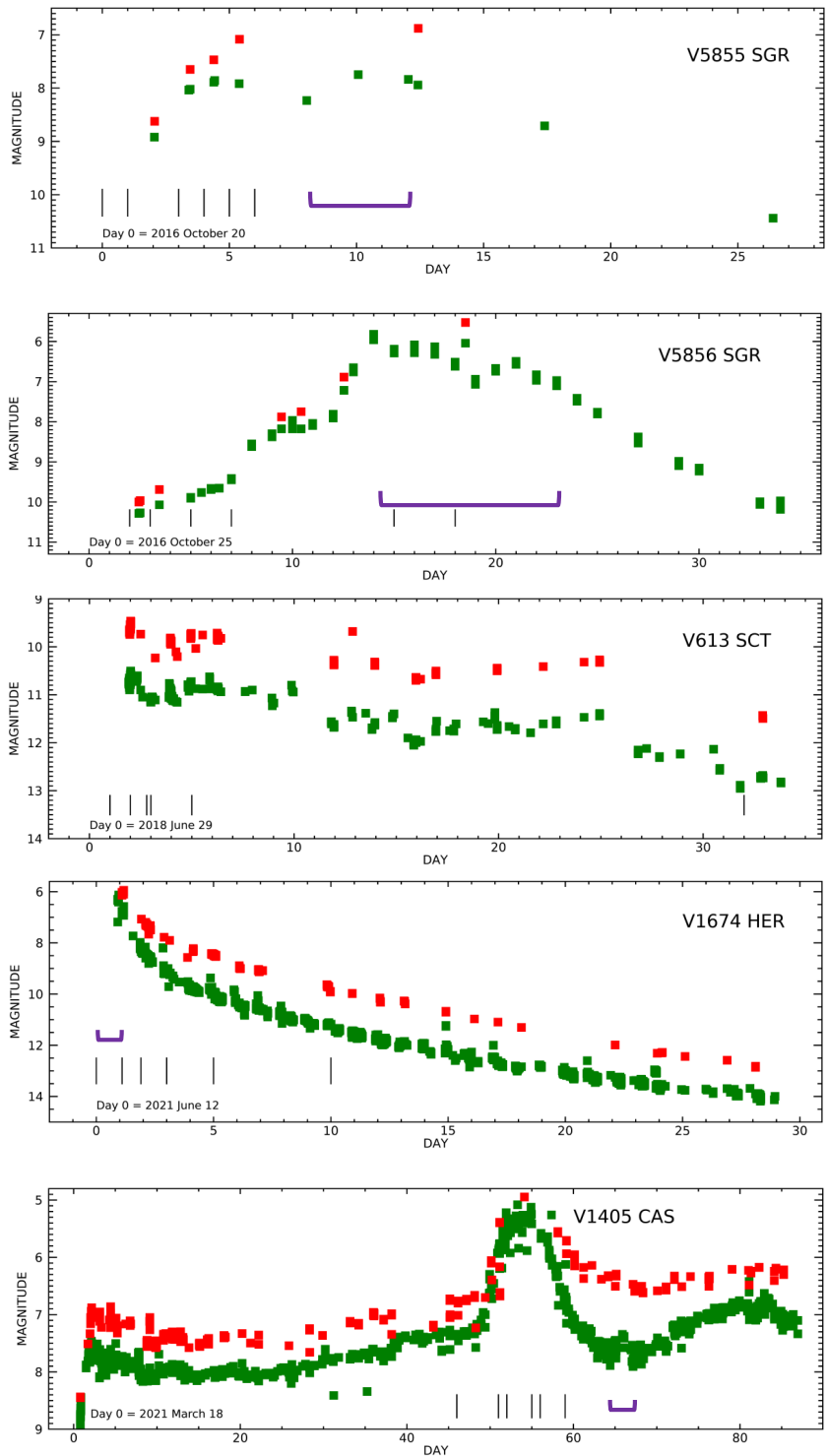

Figure 6 – Light curves for the five novae whose spectra are presented in Figs. 1-5, taken from AAVSO observations. Visible measurements are displayed in green, with r-band brightness denoted by red squares. Vertical lines mark the dates on which the ARAS spectra were obtained that are shown in Figs. 1-5. Purple horizontal brackets denote the time interval when γ-ray emission was detected by the Fermi LAT.